\documentclass[aps,showpacs,preprintnumbers,amsmath,amssymb]{revtex4}
\usepackage{graphicx}% Include figure files
\usepackage{dcolumn}% Align table columns on decimal point
\usepackage{bm}% bold math
\begin{document}

\title{The Second Law For the Transitions Between the Non-equilibrium 
Steady States}
\author{G. Baris Ba\u{g}c\i$^{1,}$}\email{baris.bagci@ege.edu.tr}
\author{Ugur Tirnakli$^1$}
%\email{ugur.tirnakli@ege.edu.tr}
\author{Juergen Kurths$^{2,3}$}
\affiliation {$^1$Department of Physics, Faculty of Science, Ege University, 
35100 Izmir, Turkey\\
$^2$Potsdam Institute for Climate Impact Research, P.O. Box 60 12 03, 
14412 Potsdam, Germany \\ 
$^3$Department of Physics, Humboldt University Berlin, Newtonstr. 15, 
12489 Berlin, Germany}

\pagenumbering{arabic}

\begin{abstract}
We show that the system entropy change for the
transitions between non-equilibrium steady states arbitrarily far
from equilibrium for any constituting process is given by the
relative entropy of the distributions of these steady states. This
expression is then shown to relate to the dissipation relations of
both Vaikuntanathan and Jarzynski [\textit{EPL} \textbf{87}, 60005
(2009)] and Kawai, Parrondo and Van den Broeck [\textit{Phys. Rev.
Lett.} \textbf{98}, 080602 (2007)] in the case of
energy-conserving driving.
%\pacs{05.20.-y}
%\pacs{05.40.-a}
\end{abstract}

\newpage \setcounter{page}{1}

\maketitle

%%%%%%%%%%%%%%%%%%%%%%%%%%%%%%%%%%%%%%%%%
\section{Introduction}\label{intro}
%%%%%%%%%%%%%%%%%%%%%%%%%%%%%%%%%%%%%%%%%
Recently there have been many novel approaches in understanding
physical systems driven away from equilibrium. Among such
approaches we emphasize the work theorems of Jarzynski \cite{Jar1}
and Crooks \cite{Crooks1}, steady-state thermodynamics
\cite{Oono,Hatano} and stochastic thermodynamics
\cite{Sekimoto,Seifert1,Ge,Qian}. In this very same context, a new
and profound understanding of the second law of thermodynamics has
been possible through its relation to the dissipation in phase
space \cite{Kawai1,Kawai2,Horowitz}, the relation between
dissipation and lag in irreversible processes
\cite{Vaikuntanathan}, and the Landauer principle and the second
law \cite{Broeck}.

An important progress concerning the second law has recently been
made by Esposito and Van den Broeck
\cite{Esposito1,Esposito2,Esposito3} where they have shown that
the total entropy production stems from adiabatic and
non-adiabatic contributions. Although both of these entropy
productions are non-negative, they are very different in nature,
since these two distinct contributions represent two different
manners of driving a system out of equilibrium. The adiabatic
entropy production in a physical system occurs when the system is
driven through non-equilibrium boundary conditions. It should be
noted that this use of the term adiabatic does not refer to
the absence of heat exchange, but to the instantaneous relaxation
to the steady state as remarked in \cite{Esposito2}. On the other
hand, the non-adiabatic entropy production stems from the external
driving with non-energy-conserving dynamics. Esposito and Van den
Broeck have further shown that the system entropy production is
the non-adiabatic entropy production minus the excess term
\cite{Esposito2}.

In a related context, Speck and Seifert \cite{Seifert2} have
recently shown that the equilibrium form of the
fluctuation-dissipation theorem (FDT) can be used for a colloidal
particle in a periodic potential if one measures the velocity with
respect to the local mean velocity. We also note that a modified
fluctuation-dissipation relation for a non-equilibrium steady
state has recently been experimentally verified \cite{Gomez}.
Speck and Seifert end their discussion by asking whether some
other non-velocity-like concepts can be used in their equilibrium
form when one studies non-equilibrium steady states.

Considering the colloidal particle in a periodic potential studied
by Speck and Seifert \cite{Seifert2}, we note that this model is
particular in that its average excess heat is zero. Motivated by
this observation, we aim to show that in this contribution the
equilibrium form of the second law can be used for the transitions
between the non-equilibrium steady states as Speck and Seifert
have shown the restoration of the equilibrium FDT for the same
class of transitions. However, it is worth remarking that our
treatment is general and not limited only to the Brownian
particles in periodic potential.

The paper is organized as follows: In the next section, we derive
the formula for the entropy change associated with the transitions
between non-equilibrium steady states arbitrarily far from
equilibrium. Next, examples are provided to clarify the use of the
previously derived main equation. The relation of the present work
to the entropy change for the transitions between the equilibrium
states and to the previous works derived in the case of
energy-conserving dynamics is then discussed. Finally, concluding
remarks are presented.

%%%%%%%%%%%%%%%%%%%%%%%%%%%%%%%%%%%%%%%%%%%%%
\section{Theory}
%%%%%%%%%%%%%%%%%%%%%%%%%%%%%%%%%%%%%%%%%%%%%

In stochastic thermodynamics, the entropy of a physical system is
given by its time-dependent Shannon entropy

\begin{equation}\label{systementropy}
S\left( t\right) =-\sum\limits_{m}p_{m}\left( t\right) \ln
p_{m}\left( t\right)
\end{equation}

\noindent where we set the Boltzmann constant equal to unity. The
general framework can then be formulated by assuming a Markovian
dynamics together with a local detailed balance condition
\cite{Esposito1,Esposito2,Esposito3}. However, in the present
work, we focus only on the non-adiabatic contribution arising from
the transitions between the non-equilibrium steady states.
Therefore, we consider a system initially in a normalized steady
state $p_{m}^{st}\left( 0\right)$ corresponding to the initial
value of the control parameter $\lambda _{t_{i}}$. The external
driving is represented as usual by the change of the control
parameter $\lambda _{t_{i}}$ to its final value $\lambda
_{t_{f}}$. Assuming that a steady state is formed for fixed values
of the control parameter after an asymptotically long time, a new
normalized steady state distribution $p_{m}^{st}\left( T\right)$
is reached after a long time $T$. For the transitions between the
non-equilibrium steady states \cite{Esposito2}, the change in the
system entropy then reads

\begin{equation}\label{systementropychange}
\Delta S=-\sum\limits_{m}p_{m}^{st}\left( T\right) \ln
p_{m}^{st}\left( T\right) +\sum\limits_{m}p_{m}^{st}\left(
0\right) \ln p_{m}^{st}\left( 0\right)
\end{equation}

\noindent and the following relation is satisfied

\begin{equation}\label{nonadiabaticchange}
\Delta S_{na}-\Delta S_{ex}=\Delta S
\end{equation}

\noindent where the excess entropy change is

\begin{equation}\label{excesschange}
\Delta S_{ex}=\sum\limits_{m}\left( p_{m}^{st}\left( T\right)
-p_{m}^{st}\left( 0\right) \right) \ln p_{m}^{st}\left( T\right).
\end{equation}

The non-adiabatic entropy production is due to the external
driving with non-energy-conserving dynamics, and the excess
contribution is what is left of the total dissipation once the
adiabatic entropy production is used to maintain the
non-equilibrium steady state. It is seen from
Eq.~(\ref{nonadiabaticchange}) that the system entropy change is
not equal to the non-adiabatic entropy change due to the excess
entropy contribution. For a Brownian particle in a periodic
potential, the excess entropy change is zero, since the steady
states in this model also correspond to the equilibrium
distributions once they are reached \cite{Seifert2,Esposito3}.
However, the excess change is not zero in general. Motivated by
the particular model studied by Speck and Seifert \cite{Seifert2},
we now generally include the excess entropy term in the system
entropy change for the transitions between the non-equilibrium
steady states. This incorporation is necessary in order to relate
our results to the recent formulae derived for the transitions
between equilibrium states, since the latter are formulated in
terms of the system entropy change only. Note that the inclusion
of the excess term into the system entropy change also makes the
latter equal to the non-adiabatic entropy change as can be seen
from Eq.~(\ref{nonadiabaticchange}). To include the excess entropy
term, we treat as if it is zero and consider the result of this
equality as a condition to be satisfied by the system entropy
change which reads

\begin{equation}\label{renormalization}
\sum\limits_{m} p_{m}^{st}\left( T\right)\ln p_{m}^{st}\left(
T\right) =\sum\limits_{m} p_{m}^{st}\left( 0\right)  \ln
p_{m}^{st}\left( T\right)
\end{equation}

\noindent as can be easily seen from Eq.~(\ref{excesschange}).
Incorporating Eq.~(\ref{renormalization}) into the change of the
system entropy given by Eq.~(\ref{systementropychange}), we obtain

\begin{equation}\label{step1}
\Delta S^{st}=-\sum\limits_{m}p_{m}^{st}\left( 0\right) \ln
p_{m}^{st}\left( T\right) +\sum\limits_{m}p_{m}^{st}\left(
0\right) \ln p_{m}^{st}\left( 0\right)
\end{equation}

\noindent which can be rewritten as

\begin{equation}\label{step2}
\Delta S^{st}=\sum\limits_{m}p_{m}^{st}\left( 0\right) \ln \left( \frac{%
p_{m}^{st}\left( 0\right) }{p_{m}^{st}\left( T\right) }\right),
\end{equation}

\noindent where we have denoted the system entropy change $\Delta
S$ including the excess term by $\Delta S^{st}$. Before proceeding
further, one can check whether considering excess entropy term as
zero in the form of a constraint and its inclusion in the system
entropy change resulted any mathematical discrepancies. Note that
one has initially the
non-adiabatic entropy change equal to 
$\Delta S_{na}=\sum\limits_{m}p_{m}^{st}\left( 0\right) \ln \left( \frac{%
p_{m}^{st}\left( 0\right) }{p_{m}^{st}\left( T\right) }\right)$ as
can be seen from Eqs. (2)-(4). Since we have obtained exactly the
same expression in Eq.~(\ref{step2}) for the system entropy
change, it implies that the excess term is well incorporated into
the system entropy without loss of generality.

We now note that the right hand side of Eq.~(\ref{step2}) is the
non-negative relative entropy expression 
$D\left[ p^{st}\left( 0\right) \Vert p^{st}\left( T\right) %
\right]$ (also known as Kullback-Leibler distance) between the
steady state distributions \cite{Cover} i.e.,

\begin{equation}\label{main}
\Delta S^{st}=D\left[ p^{st}\left( 0\right) \Vert p^{st}\left( T\right) %
\right]\geq 0 \,\, .
\end{equation}

This is our main result, which relates the system (or,
equivalently, non-adiabatic) entropy change due to the transitions
between non-equilibrium steady states to the relative entropy of
the initial and final steady state distributions. It shows that
the equilibrium form of the second law is preserved if the excess
entropy change is incorporated into the system entropy change.
This can also be seen by inspecting the second law derived by
Hatano and Sasa \cite{Hatano} i.e., $T\Delta S\geq -Q_{ex}$. This
expression of the second law immediately gives our main result
Eq.~(\ref{main}), once the excess heat $Q_{ex}$, being equal to
$T\Delta S_{ex}=Q_{ex}$, is incorporated into the system entropy
so that one now has $\Delta S^{st}\geq 0$, since $\Delta S+\Delta
S_{ex}=\Delta S^{st}$. This is plausible, since the Hatano-Sasa
form of the second law is derived for a system under
nonconservative driving in contact with a single reservoir, and
the excess entropy change is the same as the excess heat divided
by the temperature of the reservoir under this condition.

On the other hand, it is worth noting that the result in
Eq.~(\ref{main}) is more general than the one derived by Hatano
and Sasa \cite{Hatano}. The reason is that the system entropy
change now possesses all the information of the non-adiabatic
entropy change as a result of incorporating the excess entropy
change as can be seen from Eq.~(\ref{nonadiabaticchange}). It is
well-known that the non-adiabatic entropy change is related to the
system properties, and independent of the constituting process
\cite{Esposito3}. Therefore, our main result too, being exactly at
the same level of description as the non-adiabatic entropy change,
is independent of the process, which generates the steady state
conditions, be it through the time-dependent driving or multiple
reservoirs with different thermodynamic properties. However, the
Hatano-Sasa relation is valid for the transitions between
non-equilibrium steady states only under the assumption of
time-dependent driving and coupling to a single reservoir. We also
note that Eq.~(\ref{main}) is valid for any kind of transitions,
be it slow or not.

Another related issue is concerning when the equality holds for
the second law given by Eq.~(\ref{main}). In the case of the
Hatano-Sasa relation, the equality holds only for slow processes
so that one obtains $T\Delta S= -Q_{ex}$. Since the excess
contribution is incorporated into the system entropy in
Eq.~(\ref{main}), the Hatano-Sasa relation for slow processes
becomes $\Delta S^{st}= 0$. If the result of including excess
entropy change in the system entropy change would be tantamount to
rewriting the Hatano-Sasa relation, one would expect our main
result i..e, Eq.~(\ref{main}) too to be zero for slow processes.
However, being valid for arbitrary protocols and independent of
the constituting process, the second law written in terms of the
system entropy given by Eq.~(\ref{main}) is zero for two cases.
i)~The first trivial case is when the steady state distribution,
once reached, remains unchanged despite the presence of the
time-dependent driving. ii)~The second case is nontrivial and a
direct result of including the excess entropy term in the system
entropy: the entropy change given by Eq.~(\ref{main}) is zero also
when the steady state distribution does not change its form
drastically. To understand this, consider a case where
$p_{m}^{st}\left( 0\right) =\left( \frac{e^{\alpha }-1}{e^{\alpha
}}\right) e^{-\alpha m}$ and $p_{m}^{st}\left( T\right) =\left(
\frac{e^{\delta }-1}{e^{\delta }}\right) e^{-\delta m}$ are
initial and final normalized steady state distributions,
respectively. The terms $\alpha$ and $\delta$ are constants which
usually depend on the values of friction and diffusion, for
example. Enforcing Eq.~(\ref{renormalization}) i.e., incorporating
the excess entropy change into the system itself, a simple
calculation shows that $\alpha=\delta$, yielding zero system
entropy change in Eq.~(\ref{main}). This provides a new insight
into the physical meaning of slow processes regarding the
transitions between non-equilibrium steady states. In other words,
a process is slow only if the steady distribution remains the same
despite the external driving or does not change its form
drastically. One observes no entropy change in the physical system
if an initial exponential steady state distribution is preserved
exactly, or only changes its argument under the influence of
external driving or due to the coupling of the physical system to
multiple reservoirs with different thermodynamic properties.
However, a transition from an exponential steady state
distribution to a Gaussian one produces a nonzero entropy change
as we see below for the Van der Pol oscillator under the influence
of noise.

One might ask whether our central result Eq.~(\ref{main}) provides
information on the entropy change when the physical system is
acted on by an energy-conserving and conservative driving (or
considering the generality of Eq.~(\ref{main}), one can have
equilibrium states by keeping the thermodynamic properties of the
multiple reservoirs same), so that the steady states are replaced
by the equilibrium states in Eq.~(\ref{main}). In fact, all the
steps above can be repeated in their full generality only by
noting that the non-equilibrium steady states are replaced by the
equilibrium ones so that one obtains the relation 
$\Delta S^{eq}=D\left[ p^{eq}\left( 0\right) \Vert p^{eq}\left( T\right) %
\right]\geq 0$ as the second law for the transitions between
initial and final equilibrium distributions, where the equality is
satisfied for quasi-static processes.

In order to see that this is in fact the case, we consider the
non-equilibrium Landauer principle in Ref. \cite{Broeck}. This
principle in its full generality reads

\begin{equation}\label{landauer}
\beta W_{diss}-D\left[ p\left( T\right) \Vert p^{eq}\left(
T\right) \right] +D\left[ p\left( 0\right) \Vert p^{eq}\left(
0\right) \right] =\Delta _{i}S,
\end{equation}

\noindent where $\beta$ is the inverse temperature. The dissipated
work $W_{diss}$ is given by $\left\langle W\right\rangle -\Delta
F^{eq}$, where $\left\langle W\right\rangle$ and $\Delta F^{eq}$
stand for the average work and free energy difference between the
equilibrium states, respectively. In the equation above, the
relative entropy terms are separately zero if the system is both
initially and finally at equilibrium \cite{Broeck}. Moreover,
$\Delta _{i}S$ is total entropy change and becomes equal to
$\Delta S_{na}$, since the adiabatic contribution $\Delta S_{a}$
vanishes for a single reservoir with conservative driving or for
multiple reservoirs with the same thermodynamical properties due
to the local detailed balance condition
\cite{Esposito1,Esposito2,Esposito3}. Moreover, since the excess
entropy change is incorporated into the system entropy in deriving
Eq.~(\ref{main}) (see also Eq.~(\ref{nonadiabaticchange})), the
non-adiabatic entropy change $\Delta S_{na}$ is equal to the
change in the system entropy $\Delta S$. Under the influence of
conservative driving, the steady states now relax to the
equilibrium states (see e.g. the paragraph above Eq. (5) in Ref.
\cite{Tang}) so that we have

\begin{equation}\label{landauer2}
\beta W_{diss} =\Delta S^{eq}=D\left[ p^{eq}\left( 0\right) \Vert p^{eq}
\left( T\right) \right]\geq 0 \,\, .
\end{equation}

\noindent We note that $W_{diss}\geq 0$ is indeed the second law
of thermodynamics, and the dissipated work too, being explicitly
equal to $\left\langle W\right\rangle -\Delta F^{eq}$, relates
only two equilibrium states just like the relative
entropy expression $D\left[ p^{eq}\left( 0\right) \Vert p^{eq}\left( T\right) %
\right]$. Moreover, this equality also ensures that the relative
entropy formulation of the second law attains zero only for
quasistatic, reversible processes, since only then the dissipated
work vanishes. By Eq.~(\ref{landauer2}), one can further see why
our main result Eq.~(\ref{main}) has been interpreted as
preserving the equilibrium form of the second law.

Although we have only considered transitions between equilibrium
states above, one can use the non-equilibrium Landauer principle
\cite{Broeck} to include, for example, the transition from an
initial equilibrium state to a final non-equilibrium one. Hence,
one can write Eq.~(\ref{landauer}) as

\begin{equation}\label{landauer3}
\Delta S^{eq}-D\left[ p\left( T\right) \Vert p^{eq}\left( T\right)
\right] =\Delta _{i}S,
\end{equation}

\noindent where the term $D\left[ p\left( 0\right) \Vert
p^{eq}\left( 0\right) \right]$ is zero, since the system is
initially at equilibrium. One has to keep the right hand side of
the equation above as $\Delta _{i}S$, since the final state is not
at equilibrium any more. In other words, one does not consider
only the transitions between the equilibrium states, and
consequently boundary terms must be included too (see in
particular Eqs.~(37-39), and (42) in Ref. \cite{Esposito2}).
Therefore, $\Delta _{i}S$ cannot be equal to $\Delta S^{eq}$ in
general as we had in Eq.~(\ref{landauer2}). However, the total
entropy $\Delta _{i}S$ is always non-negative \cite{Broeck} i.e.,
$\Delta _{i}S\geq 0$, implying

\begin{equation}\label{vaikunathan}
\Delta S^{eq}\geq D\left[ p\left( T\right) \Vert p^{eq}\left(
T\right) \right].
\end{equation}

\noindent This expression was recently obtained by Vaikuntanathan
and Jarzynski \cite{Vaikuntanathan} for the relation between the
dissipation and the lag where they have used $\beta W_{diss}$
instead of its equal $\Delta S^{eq}$ (see Eq.~(\ref{landauer2})
above). Finally, following Ref. \cite{Vaikuntanathan} (see in
particular the section below Eq. (10) in Ref.
\cite{Vaikuntanathan}), one can write Eq.~(\ref{vaikunathan}) as

\begin{equation}\label{kawai}
\Delta S^{eq}\geq D\left[ p\left( T\right) \Vert
\widetilde{p}^{eq}\left( 0\right) \right],
\end{equation}

\noindent where tilde denotes the reverse process so that the
initial state of the system during the reverse process is the
final equilibrium state i.e., $p^{eq}\left(
T\right)=\widetilde{p}^{eq}\left( 0\right)$. This last relation
was obtained by Kawai {\it et al.} \cite{Kawai1,Kawai2}, and relates the
dissipation to the time reversal asymmetry. Eq.~(\ref{kawai}), on
the other hand, relates the change in equilibrium entropies to the
time-reversed process.

%%%%%%%%%%%%%%%%%%%%%%%%%%%%%%%%%%%%%%%%%%%%%
\section{Examples}
%%%%%%%%%%%%%%%%%%%%%%%%%%%%%%%%%%%%%%%%%%%%%

Before proceeding, we would like to note that the results of the
previous section are valid in the continuous case even though our
formalism has been discrete so far.

As an illustrative model, we first consider a driven Brownian
particle on a circle \cite{Esposito3,Seifert2}

\begin{equation}\label{driven}
\dot{x}=u_{t}+\sqrt{2D}\xi \,\, ,
\end{equation}

\noindent where $u_{t}$ is the time dependent drift and $D$ is the
time independent diffusion constant, also assuming $x\in \lbrack
0,1]$. This model represents a colloidal particle moving in a
periodic potential, and it is used to study the violation of the
fluctuation-dissipation theorem in non-equilibrium steady states
with external driving \cite{Seifert2}. Note that the force is
directly proportional to the drift term in stochastic
thermodynamics (see Eq. (15) in Ref. \cite{Esposito3}). The
stationary solution for the driven Brownian particle is equl to
unity for any value of the control parameter i.e. $p^{st}\left(
0\right) =p^{st}\left( T\right) =1$ \cite{Esposito3,Seifert2}. In
other words, once the Brownian particle has relaxed to this steady
state with $p^{st}=1$, it remains so despite any external driving.
Our main result i.e., Eq.~(\ref{main}), yields zero entropy change
for this model. Note that this result was also observed in Ref.
\cite{Esposito3} for the non-adiabatic entropy change using the
Fokker-Planck formulation of the stochastic thermodynamics once
the steady state distribution is reached, rendering the excess
entropy contribution redundant from there on for the transitions
between the non-equilibrium steady states .

A non-trivial example is the Van der Pol oscillator subject to
noise. Then the Ito-Langevin type stochastic equation reads

\begin{eqnarray}\label{vanderpol}
\dot{x} & = & v  \nonumber \\
\\
\dot{v}+\left( a+bE\right) v+x & = & \eta (t) \nonumber
\end{eqnarray}

\noindent where $a$ and $b$ are the controllable linear and fixed
nonlinear friction coefficients, respectively
\cite{Klimontovich2}. The term $E$ denotes, setting the mass and
the angular frequency equal to unity for simplicity, the energy of
the oscillation i.e., $E=\frac{1}{2}\left( v^{2}+x^{2}\right)$.
The random noise is defined to be Gaussian with the noise
intensity $\sqrt{2D}$ i.e., $\left\langle \eta (t)\right\rangle
=0,\left\langle \eta (t)\eta (t^{\prime })\right\rangle =2D\delta
\left( t-t^{\prime }\right)$. We assume that we can control the
change in the linear frictional term i.e., $a=\gamma-\lambda$ and
$\gamma\gg \frac{bE}{2}$ where $\lambda$ denotes the control
parameter and $\gamma$ is the linear friction coefficient by
default. The most general stationary solution of the noise-driven
Van der Pol oscillator then reads

\begin{equation}\label{generalss}
p^{st}\left(\lambda \right)=\exp
\left(\frac{-aE-\frac{1}{2}bE^{2}}{D}\right)
\end{equation}

\noindent apart from the appropriate normalization
\cite{Klimontovich2}.  From here on, we use the energy
representation, since it is equivalent to the phase space
integration for the simple harmonic oscillator case. We now assume
that the physical system initially described by the stationary
distribution $p ^{st}\left(0 \right)$ with zero value of the
control parameter evolves into $p ^{st}\left(T \right)$ with
$\lambda_{B}=\gamma$ through a protocol controlled by an external
agent. The initial stationary distribution corresponding to
$\lambda=0$ is given by

\begin{equation}\label{sszero}
p ^{st}\left(0 \right)=\frac{\gamma}{D}\exp \left(-\frac{\gamma
E}{D}\right) \,\, ,
\end{equation}

\noindent where we have used $\gamma\gg \frac{bE}{2}$. The final
steady state with $\lambda=\gamma$ reads

\begin{equation}\label{ssnonzero}
p ^{st}\left(T \right)=\sqrt{\frac{2b}{\pi
D}}\exp\left(-\frac{b}{2D}E^{2}\right) \,\,.
\end{equation}

\noindent The left hand side of Eq.~(\ref{renormalization}) can
now be calculated as $\left( \frac{1}{2}\ln \left( \frac{2b}{\pi
D}\right) -\frac{1}{2}\right)$, whereas the right hand side of the
same equation yields 
$\left( \frac{1}{2}\ln \left( \frac{2b}{\pi D}\right) 
-\frac{Db}{\gamma ^{2}}\right)$. 
Therefore, one explicitly obtains from Eq.~(\ref{renormalization})

\begin{equation}\label{hadibakalim}
\frac{\gamma ^{2}}{2Db}=1 \,\,.
\end{equation}

\noindent Incorporating the relation above into Eq.~(\ref{main})
is tantamount to including the excess entropy change into the
system entropy. Eq.~(\ref{main}) by itself explicitly yields

\begin{equation}\label{sol1}
\Delta S=\int_{0}^{\infty }dE\frac{\gamma }{D}\exp \left( -\frac{\gamma }{D}%
E\right) \ln \left[ \frac{\frac{\gamma }{D}\exp \left( -\frac{\gamma }{D}%
E\right) }{\sqrt{\frac{2b}{\pi D}}\exp \left( -\frac{b}{2D}E^{2}\right) }%
\right],
\end{equation}

\noindent which, after integration, becomes

\begin{equation}\label{sol2}
\Delta S=\ln \left( \sqrt{\pi \frac{\gamma ^{2}}{2Db}}\right) -1+\frac{Db}{%
\gamma ^{2}}.
\end{equation}

\noindent Including the relation given by Eq.~(\ref{hadibakalim})
which is tantamount to including the excess entropy change in the
system entropy, we obtain

\begin{equation}\label{sol3}
\Delta S^{st}=\ln \left( \sqrt{\pi}\right) -1+\frac{1}{%
2},
\end{equation}

\noindent which finally yields

\begin{equation}\label{finalresult}
\Delta S^{st} = D\left( p ^{st}\left(0 \right) \parallel p
^{st}\left(T \right)\right) \approx \bigskip 0.07 \,\, ,
\end{equation}

\noindent which is indeed greater than zero, thereby indicating
the irreversibility of the transition between these two
non-equilibrium steady states. The smallness of this value is
expected, since the departure from the compared steady state with
zero control parameter is given with respect to the steady state
corresponding to $\lambda _{B}=\gamma$, where $\gamma$ is itself
supposed to be small in all realistic cases.

%%%%%%%%%%%%%%%%%%%%%%%%%%%%%%%%%%%%%%%%%%%%%
\section{Conclusions}
%%%%%%%%%%%%%%%%%%%%%%%%%%%%%%%%%%%%%%%%%%%%%
To summarize, the system entropy change for the transitions
between non-equilibrium steady states arbitrarily far from
equilibrium is obtained in terms of the relative entropy of the
concomitant steady state distributions. This result is independent
of the constituting process in the sense that steady states can
result either due to the non-conservative driving or through the
presence of multiple reservoirs with different thermodynamic
properties. We also note that the same expression  for the
transition between non-equilibrium steady states i.e.,
Eq.~(\ref{main}), can be used for the transitions between
equilibrium states only by replacing the stationary distributions
with the corresponding equilibrium ones. Considering only the
transitions between the equilibrium states, our result given by
Eq.~(\ref{landauer2}) implies the relations obtained in Refs.
\cite{Kawai1,Kawai2,Vaikuntanathan}. However, these previous
relations considered the dissipated work as a measure of the
second law while we have related them to the entropy change
between the equilibrium states through relative entropy
expression.

It is worth noting that one should not confuse the main result of
this paper given by Eq.~(\ref{main}) with the well-known similar
expression $D\left[ p\left( t\right) \Vert p^{st}\left( \lambda\right) %
\right]$ \cite{Esposito2}. This expression can be considered as a
proof of convergence to steady state and relates the actual and
the corresponding steady state distributions, whereas
Eq.~(\ref{main}) on the other hand relates two distinct steady
states. In this context, we also note that a new approach has
recently been introduced by defining a novel state function
information free energy which also includes the adiabatic term in
non-equilibrium thermodynamics \cite{Deffner}.

Finally, we remark that the results outlined in this work can be
experimentally tested e.g., by using the Van der Pol oscillator
studied in this work, or a motor protein coupled to an
ATP-regenerating system such that the motor protein forms a
non-equilibrium steady state \cite{Liu}.

%%%%%%%%%%%%%%%%%%%%%%%%%%%%

\end{document}